\documentclass{webofc}
\usepackage[varg]{txfonts}   % Web of Conferences font
\usepackage{wrapfig}

\newcommand{\pt}{\mbox{$p_T$}\xspace}

\newcommand{\rca}{\mbox{$R_{Cu Au}$}\xspace}

\newcommand{\pp}{\mbox{$p$$+$$p$}\xspace}
\newcommand{\dau}{\mbox{$d$$+$Au}\xspace}

\newcommand{\pau}{\mbox{$p$$+$Au}\xspace}
\newcommand{\pal}{\mbox{$p$$+$Al}\xspace}
\newcommand{\ca}{\mbox{Cu$+$Au}\xspace}
\newcommand{\ha}{\mbox{$^3$He$+$Au}\xspace}

\begin{document}
\title{$\phi$ Meson Measurements at RHIC with the PHENIX Detector}

\author{\firstname{Murad} \lastname{Sarsour}\inst{1}\fnsep\thanks{\email{msar@gsu.edu}}
on behalf of the PHENIX Collaboration }

\institute{Georgia State University, Atlanta, Georgia 30303, USA}

\abstract{%
 The measurement of $\phi$ mesons provides a unique and complementary method for
 exploring properties of the hot and dense medium created in the relativistic
 heavy ion collisions. It has a relatively small hadronic interaction cross
 section and is sensitive to the increase of strangeness (strangeness
 enhancement), a phenomenon associated with soft particles in bulk matter.
 Measurements in the dilepton channels are especially interesting since leptons
 interact only electromagnetically, thus carrying the information from their
 production phase directly to the detector. Measurements in different
 nucleus-nucleus collisions allow us to perform a systematic study of the
 nuclear medium effects on $\phi$ meson production. The PHENIX detector provides
 the capabilities to measure the $\phi$ meson production in a wide range of transverse
 momentum and rapidity to study these effects.

 In this proceeding, we present measurements of the $\phi$ mesons in a variety
 of collision systems at $\sqrt{s_{NN}}$ = 200 GeV.
 In case of small systems, the data are compared with AMPT calculations to study
 the various cold nuclear medium effects involved in $\phi$ meson production.
}
\maketitle
\section{Introduction}
\label{intro}
  Ever since RHIC announced the discovery of the hot and dense state of strongly
 interacting matter called Quark-Gluon Plasma (QGP)~\cite{NPA.757.184,NPA.757.102,
EurPhysJ.Plus.131.70}
 the main objective has been to quantify its properties. This is accomplished by
 looking at as many observables as possible, such as the nuclear modification of
 $\phi$ meson production in the QGP environment. $\phi$ meson is an excellent probe
 for studying QGP in heavy ion collisions because it is sensitive to several aspects
 of the collision, including modifications of strangeness production in bulk matter.
 Owing to its small inelastic cross section for interaction with nonstrange hadrons,
 the $\phi$ meson is less affected by late hadronic rescattering and may reflect the
 initial evolution of the system. Being composed of a nearly pure strange antistrange
 ($s\bar{s}$) state, the $\phi$ meson puts additional constraints on models of quark
 recombination in the QGP.
 However to gauge the QGP related modifications, we need to know the $\phi$ meson
 production in \pp collisions as a baseline. There are additional effects from the
 nuclear medium itself, cold nuclear matter effects (CNM), and they are accessed by
 studying $\phi$ meson production in small system collisions (e.g. \dau).

 The lepton decay channel is of particular interest because of the absence of strong
 interactions between muons and the surrounding hot hadronic matter. At forward
 rapidity, it allows us to study the rapidity dependence of $\phi$ meson production
 and especially in asymmetric heavy-ion collisions provides the means for accessing
 different mixtures of initial and final state effects.

 In \pp collisions, the $\phi$ meson production is important to study because
 it may have similar production mechanisms to other onia like $J/\psi$ and
 $\Upsilon$. In addition, it provides a tool to study effects that scale with mass
 (e.g. collective effects) since it is the heaviest easily accessible meson made of
 light quarks.

\section{Experimental Setup}
\label{expt} 

 The PHENIX detector~\cite{NIMA.499.469} has a high rate capability utilizing a fast DAQ
 and specialized triggers, high granularity detectors, and good mass resolution and
 particle ID. Detection of $e^\pm$ utilizes the finely grained electromagnetic calorimeter
 (EMCal), the drift chamber and ring imaging Cherenkov detector, while the $\mu^\pm$
 are detected by the forward spectrometers consisting of Muon ID and Muon Tracker.
 The very forward beam-beam counters (BBC) are used to determine the collision vertex
 position and time, the beam luminosity and form a minimum bias trigger.

   The PHENIX collaboration collected data from a variety of collisions systems provided
 by the Relativistic Heavy Ion Collider (RHIC) at Brookhaven National Laboratory (BNL).
 Results from some of these data sets are presented in this proceeding.

\section{Results}
\label{res}

 PHENIX has measured the production of $\phi$ meson in \pp collisions at
 $\sqrt{s}$ = 200 GeV over a wide range of \pt and
 rapidity~\cite{PRD.83.052004,PRD.90.052002}, which is a very important
 resource for validating the phenomenological models of strangeness production. In
 addition, these measurements provide baseline for studying cold and hot nuclear
 modifications.

 To gain insight into nuclear medium effects and particle production mechanisms
 in $A$+$B$ collisions, the ratio of the $\phi$ meson yields in $A$+$B$ collisions to
 \pp collisions scaled by the number of nucleon-nucleon collisions in the $A$+$B$
 system, $N_{\rm coll}$~\cite{PRC.87.034904}, is calculated as: 
\begin{equation}\label{eqn:rdau}
  R_{AB} = \frac{d^2N_{AB}/dydp_T}{N_{\rm coll}\times d^2N_{pp}/dydp_T},
\end{equation}

where $d^2 N_{AB}/dydp_{T}$ is the per-event yield of particle 
production in heavy ion collisions and $d^2 N_{pp}/dydp_{T}$ is the per-event 
yield of the same process in \pp collisions. The $p$$+$$p$ invariant yield 
used in the $A$+$B$ calculation for the $l^{+}l^{-}$~decay channel is the 
$p$$+$$p$ differential cross section~\cite{PRD.90.052002} divided by the
$p$$+$$p$ total cross section, 42.2~mb.

Measurements of $\phi$ meson $R_{AB}$ at midrapidity in heavy ion collisions
show that $\phi$ meson exhibits a different suppression pattern compared to lighter
mesons ($\pi^0$ and $\eta$) and baryons (protons and antiprotons)~\cite{PRC.83.024909}.
For all centralities, $\phi$ meson is less suppressed than $\pi^0$ and $\eta$ in
the intermediate \pt range (2--5 GeV/c), whereas $\phi$, $\pi^0$ and
$\eta$ show similar suppression at higher \pt. 
However, $\phi$ meson $R_{AB}$ at forward and backward
rapidities~\cite{PRC.92.044909} show that $\phi$ meson yields in most 
central \dau collisions are
suppressed at low-\pt ($<$ 2 GeV/c) in the $d$-going direction and strongly enhanced in
Au-going direction in the intermediate \pt range (2--3 GeV/c). This modification decreases
as we move towards peripheral collisions. No modification of $\phi$ meson production is
observed in the most peripheral \dau collisions. $\phi$
meson $R_{d{\rm Au}}$ as a function of $p_T$ was measured for different
centralities and a nuclear modification very similar to that of
the heavy flavor muons is observed~\cite{PRL.112.252301}. 
 The $\phi$ meson enhancement in the Au-going direction and the suppression in
 the $d$-going direction are consistent with what is observed by ALICE in $p$$+$Pb
 collisions at $\sqrt{s_{NN}}$=5.02~TeV in $-4.46<y<-2.96$ and
 $2.03<y<3.53$~\cite{PLB.768.203}.

 The nuclear-modification factor is also studied in \ca collisions, \rca, at
 $\sqrt{s_{NN}}$ = 200 GeV to evaluate the effects of hot and cold nuclear matter on
 $\phi$ meson production. Although statistically limited,
 we observed a dependence of \rca on both centrality and rapidity with a
 similar trend to that observed in \dau collisions~\cite{PRC.92.044909}.

%\begin{figure}[h]
\begin{wrapfigure}{r}{0.5\textwidth}
  \begin{center}
    \includegraphics[width=0.5\textwidth]{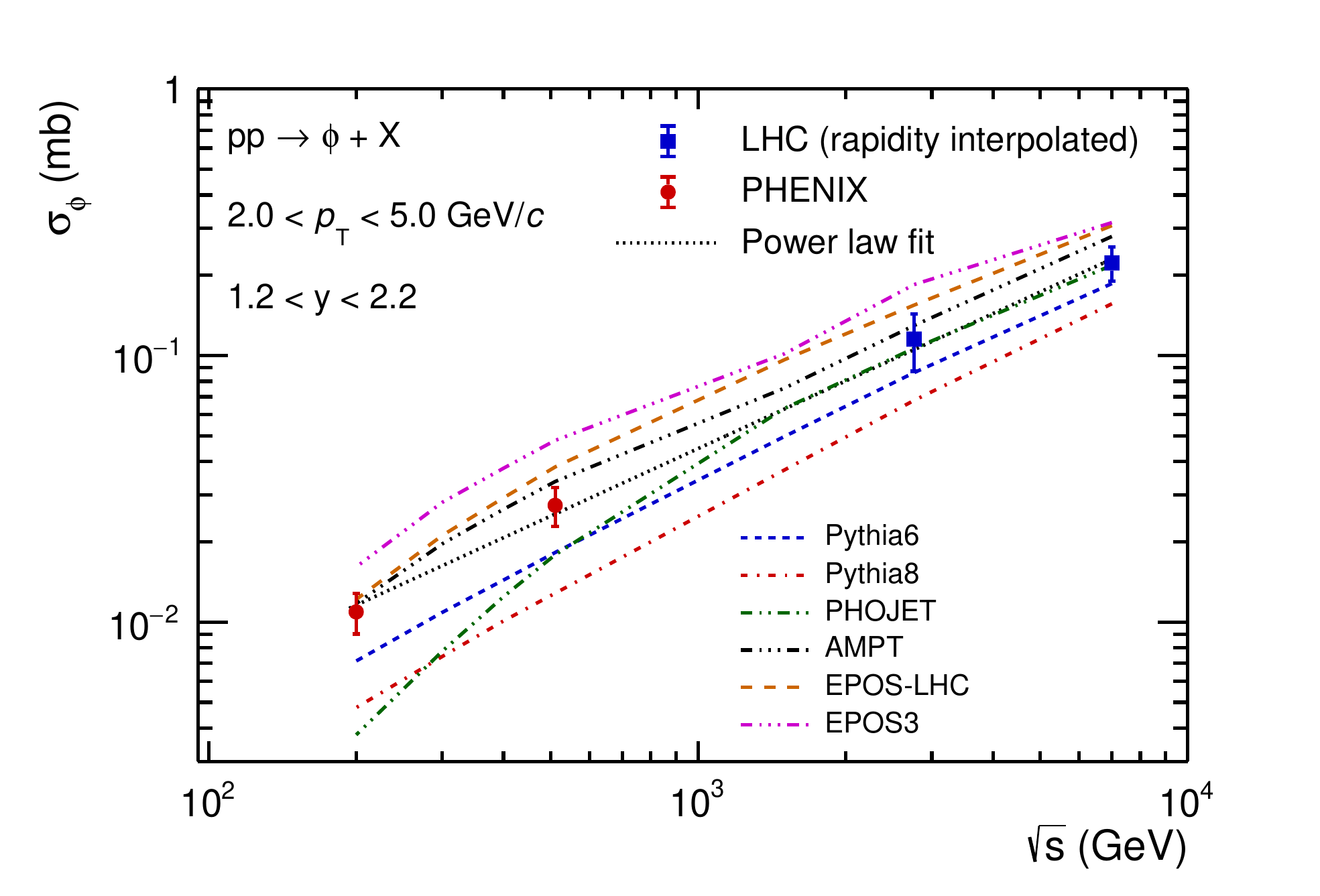}
  \end{center}
  \caption{$\phi$ meson production cross section in $1.2 < y < 2.2$ and
    $2 < p_T < 5$ GeV/c in \pp collisions versus the center-of-mass energy,
    $\sqrt{s}$, compared to different model predictions.}% The LHC data
   % points are interpolated at the PHENIX forward rapidity~\cite{arXiv.1710.01656v1}.}
  \label{fig:phisig510}
\end{wrapfigure}
%\end{figure}
 PHENIX measured the differential cross section of $\phi$ meson at forward and backward
 rapidities in \pp collisions at $\sqrt{s}=510$ GeV~\cite{arXiv.1710.01656v1}.
 The integrated cross section over the measured \pt range, 2.79 $\pm$ 0.20 (stat)
 $\pm$ 0.17 (syst) $\pm$ 0.34 (norm) $\times 10^{-2}$ mb, from this measurement and
 those from PHENIX at $\sqrt{s}=200$ GeV~\cite{PRD.90.052002} and LHC at $\sqrt{s}=$
 2.76 and 7.0 TeV~\cite{PLB.703.267,PLB.710.557,EurPhysJC.72.2183,PLB.768.203} are
 used to study the energy dependence of $\phi$ meson production at forward
 rapidity, $1.2<y<2.2$, in \pp collisions, as shown in Fig.~\ref{fig:phisig510}.
 Comparisons to PYTHIA6~\cite{JHighEnergyPhys.05.026}, PYTHIA8~\cite{ComputPhysCommun.191.159},
 PHOJET~\cite{hep-ph.9803437}, AMPT~\cite{PRC.72.064901}, EPOS3~\cite{PRC.89.064903} and
 EPOS-LHC~\cite{PRC.92.034906} model predictions show that these models exhibit similar
 energy dependence trend to that of the data. Further studies with EPOS3 model show
 that the addition of the hydrodynamic evolution of the system induces an enhancement
 of $\phi$ meson production at the LHC energies at intermediate \pt, which hints
 at collective effects, however, no effect is observed at RHIC energies~\cite{arXiv.1710.01656v1}.

  PHENIX collected data from \ha, \pau and \pal collisions in 2014 and 2015.
 In addition to $d$$+$Au data set collected in 2008~\cite{PRC.83.024909}, these data
 allow to carry out a systematic study of the various cold nuclear matter effects
 included in models like AMPT and EPOS.
\begin{figure}[h]
\centering
\includegraphics[width=7cm,clip]{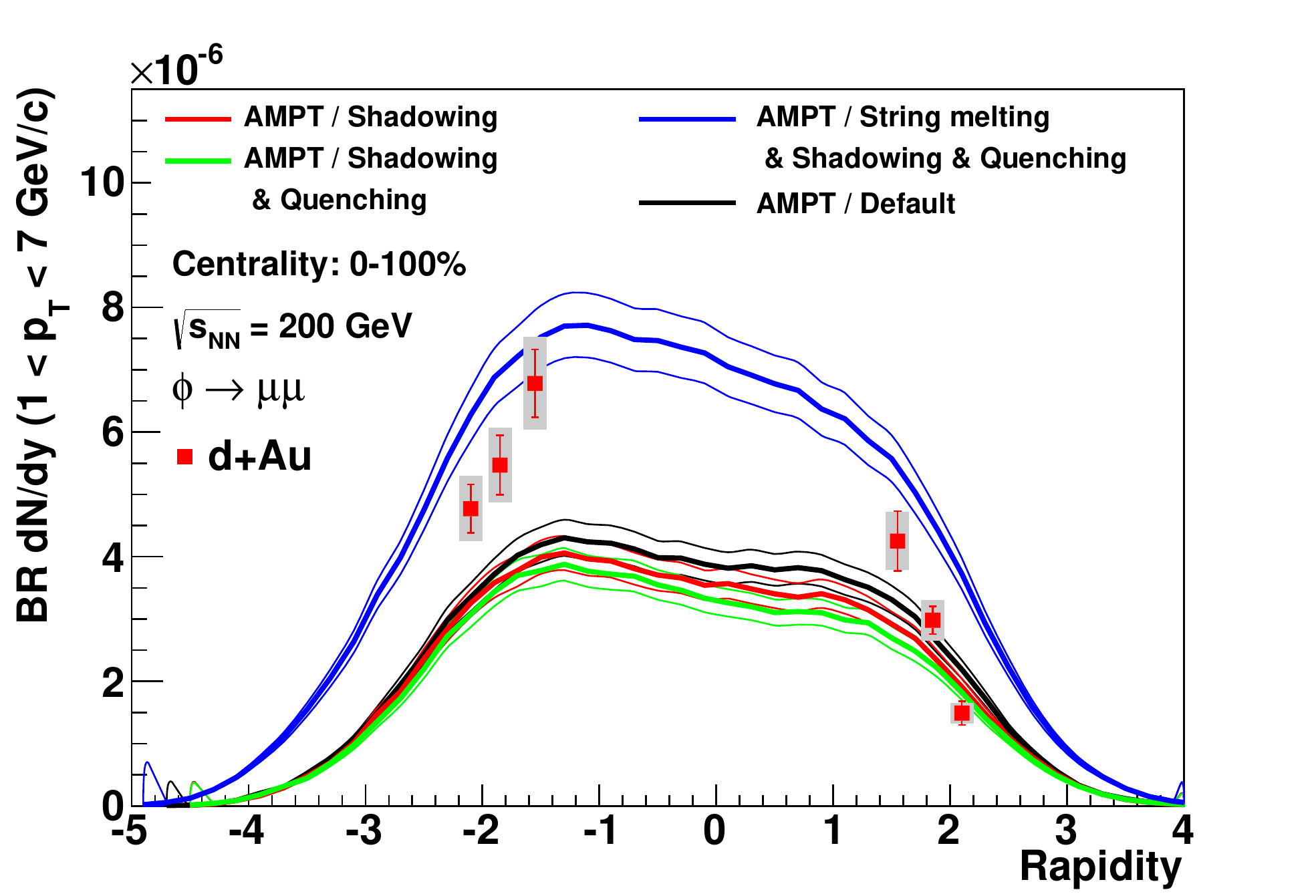}
\includegraphics[width=7cm,clip]{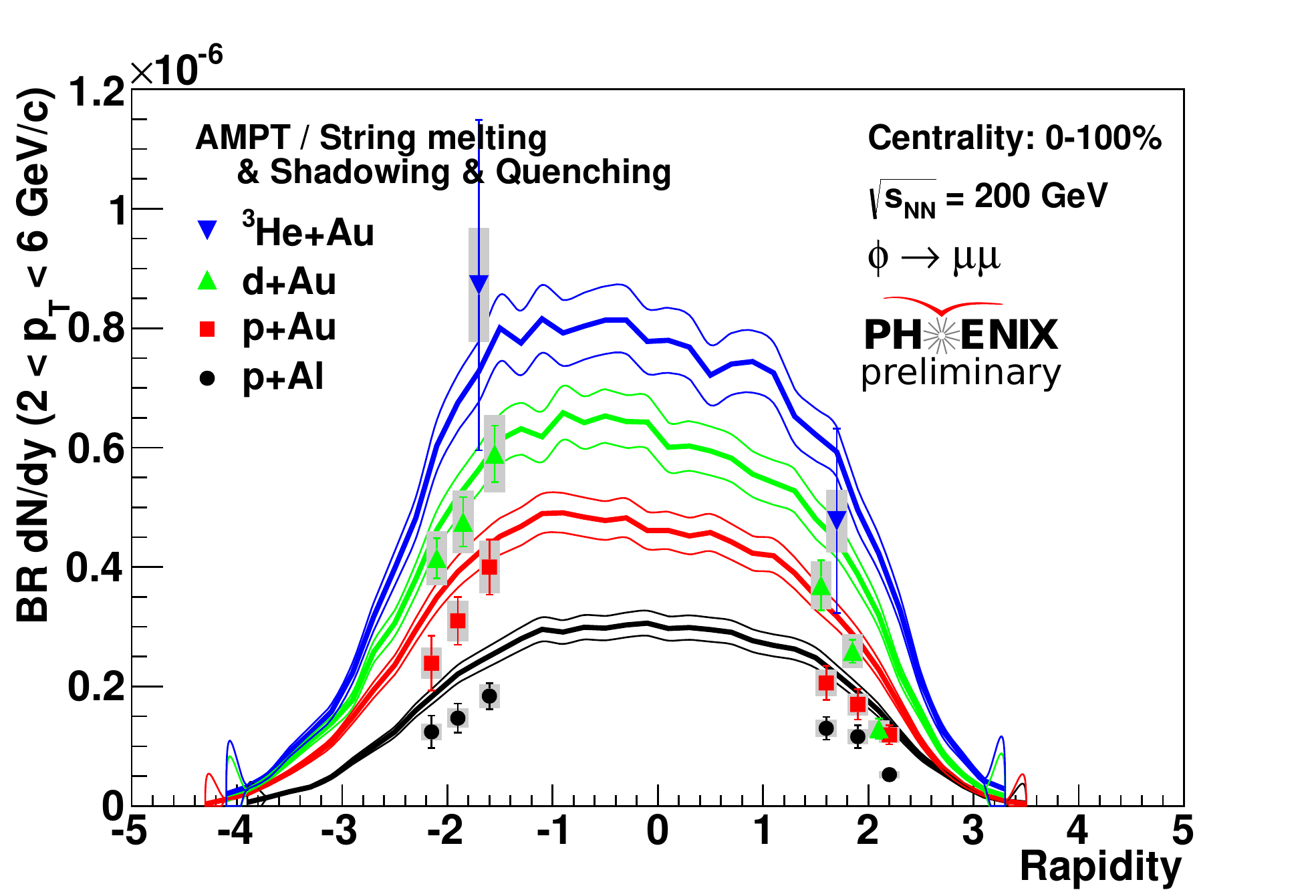}
\caption{Left: $\phi$ meson invarient yield compared with AMPT model predictions
 with different combinations of nuclear medium effects. Right: $\phi$ meson invarient
 yields from \ha, $d$$+$Au, \pau and \pal collisions compared with AMPT model
 predictions that include string melting, shadowing and quenching~\cite{PRC.72.064901}.}
\label{fig-5}       % Give a unique label
\end{figure}
 The left panel of Fig.~\ref{fig-5} shows AMPT model predictions with different combination of effects included against $\phi$ meson invariant yield in $d$$+$Au collisions. The right panel shows the AMPT model predictions, with the string melting, shadowing and quenching turned on, compared to the data.

\section{Summary}
\label{summ}

The PHENIX collaboration measured $\phi$ meson production, 
over a wide \pt range in the forward and backward rapidities 
in a variety of collision systems to study cold and hot nuclear matter effects.
In \dau collisions, we observed an enhancement (suppression) of $\phi$
 meson production in the backward (forward) rapidity region in most central collisions.
 Similar behavior was previously observed for inclusive charged hadrons and
 open heavy flavor muons which may suggest similar cold nuclear matter effects.

 The PHENIX collaboration also measured $\phi$ meson production in \pp collisions
 at $\sqrt{s}=510$ GeV at backward and forward rapidities, $\sigma_\phi$ = 2.79 $\pm$
 0.20 (stat) $\pm$ 0.17 (syst) $\pm$ 0.34 (norm) $\times 10^{-2}$ mb. Combined with
 previous PHENIX and LHC measurements, this measurement allowed studying the energy
 dependence of $\phi$ meson production in \pp collisions over a wide energy range.
 Comparisons with EPOS3 model showed a hint of collective effects in \pp at LHC
 energies and no effect at RHIC energies.

 New data sets from \pau and \pal collisions collected in 2015 allow $\phi$
 measurement at backward and forward rapidities in less complicated \pau and \pal
 systems. The wealth of small system data sets (\ha, \dau, \pau and \pal) along
 with the introduction of the forward vertex detector in 2015 data sets will
 allow studying the different CNM effects using models like AMPT and EPOS.

% BibTeX or Biber users please use (the style is already called in the class, ensure that the "woc.bst" style is in your local directory)
% \bibliography{name or your bibliography database}

\end{document}